\documentclass[]{spie}  %>>> use for US letter paper
\usepackage[]{graphicx}

\title{The High Energy cosmic-Radiation Detection (HERD) Facility onboard China's Future Space Station}

\author{S.N. Zhang\supit{a}, O. Adriani\supit{p,q}, S. Albergo\supit{t}, G. Ambrosi\supit{r},
Q. An\supit{e}, T.W. Bao\supit{a},\\
R. Battiston\supit{r,s},
X.J. Bi\supit{a}, Z. Cao\supit{a}, J.Y. Chai\supit{a}, J. Chang\supit{c},
G.M. Chen\supit{a}, Y. Chen\supit{f}, X.H. Cui\supit{i}, Z.G. Dai\supit{f},
R. D'Alessandro\supit{p,q}, Y.W. Dong\supit{a}, Y.Z. Fan\supit{c}, C.Q. Feng\supit{e},
H. Feng\supit{h}, Z.Y. Feng\supit{a},\\
X.H. Gao\supit{d}, F. Gargano\supit{o},
N. Giglietto\supit{o}, Q.B. Gou\supit{a}, Y.Q. Guo\supit{a}, B.L. Hu\supit{d},
H.B. Hu\supit{a}, H.H. He\supit{a}, G.S. Huang\supit{e}, J. Huang\supit{a},
Y.F. Huang\supit{f}, H. Li \supit{a}, L. Li\supit{a}, Y.G. Li\supit{a}, Z. Li\supit{g},
E.W. Liang\supit{l}, H. Liu\supit{a}, J.B. Liu\supit{e}, J.T. Liu\supit{a}, S.B. Liu\supit{e}, S.M. Liu\supit{c}, X. Liu\supit{a},  J.G. Lu\supit{b},
M.N. Mazziotta\supit{o}, N. Mori\supit{p,q},\\
S. Orsi\supit{v}, M. Pearce\supit{w},
M. Pohl\supit{u}, Z. Quan\supit{a}, F. Ryde\supit{w}, H.L. Shi\supit{a},
P. Spillantini\supit{p,q}, M. Su\supit{x,y},\\
J.C. Sun\supit{a}, X.L. Sun\supit{b}, Z.C. Tang\supit{a},
R. Walter\supit{v}, J.C. Wang\supit{j}, J.M. Wang\supit{a}, L. Wang\supit{d},\\
R.J. Wang\supit{a}, X.L Wang\supit{e}, X.Y. Wang\supit{f}, Z.G. Wang\supit{b},
D.M. Wei\supit{c},  B.B. Wu\supit{a}, J. Wu\supit{n}, X. Wu\supit{u}, X.F. Wu\supit{c},
J.Q. Xia\supit{a}, H.L. Xiao\supit{a}, H.H. Xu\supit{a}, M. Xu\supit{a}, Z.Z. Xu\supit{e},
H.R. Yan\supit{g}, P.F. Yin\supit{a},\\
Y.W. Yu\supit{m}, Q. Yuan\supit{a},
M. Zha\supit{a}, L. Zhang\supit{k}, L. Zhang\supit{a}, L.Y. Zhang\supit{a}, Y. Zhang\supit{a}, Y.J. Zhang\supit{a}, Y.L. Zhang\supit{e}, Z.G. Zhao\supit{e}
\skiplinehalf
\supit{a}Key Laboratory of Particle Astrophysics, Institute of High Energy Physics, Chinese Academy of Sciences, Beijing, China; \\
\supit{b}Center of Experimental Physics, Institute of High Energy Physics, Chinese Academy of Sciences, Beijing, China; \\
\supit{c}Purple Mountain Observatory, Chinese Academy of Sciences, Nanjing, China;\\
\supit{d}Xi'an Institute of Optics and Precision Mechanics, Chinese Academy of Sciences, Xi'an, China;\\
\supit{e}Department of Modern Physics, University of Science and Technology of China, Hefei, China;\\
\supit{f}School of Astronomy and Space Science, Nanjing University, Nanjing, China;\\
\supit{g}Department of Astronomy, Peking University, Beijing, China;\\
\supit{h}Department of Engineering Physics, Tsinghua University, Beijing, China\\
\supit{i}National Astronomical Observatories, Chinese Academy of Sciences, Beijing, China;\\
\supit{j}Yunnan Astronomical Observatory, Chinese Academy of Sciences, Kunming, China;\\
\supit{k}Department of Astronomy, Yunan University, Kunming, China;\\
\supit{l}Department of Physics, Guangxi University, Nanning, China;\\
\supit{m}Institute of Astrophysics, Central China Normal University, Wuhan, China;\\
\supit{n}Department of Physics, China University of Geosciences, Wuhan, China;\\
\supit{o}Istituto Nazionale di Fisica Nucleare, Bari, Italy;\\
\supit{p}INFN Sezione di Firenze, Florence, Italy;\\
\supit{q}Department of Physics and Astronomy, University of Florence, Florence, Italy;\\
\supit{r}INFN Sezione di Perugia, Perugia, Italy; \\
\supit{s}INFN-TIFPA and Universit\`{c} di Trento, Trento, Italy;\\
\supit{t}INFN Sezione di Catania and Universit\`{c} di Catania, Catania, Italy;\\
\supit{u}DPNC, University of Geneva, Geneva, Switzerland\\
\supit{v}ISDC Data Centre for Astrophysics, University of Geneva, Geneva, Switzerland\\
\supit{w}Department of Physics, Royal Institute of Technology (KTH), Stockholm, Sweden ;\\
\supit{x}Department of Physics, and Kavli Institute for Astrophysics and Space Research, Massachusetts Institute of Technology, Cambridge, USA\\
\supit{y}Institute for Theory and Computation, Harvard-Smithsonian Center for Astrophysics, Cambridge, USA\\
}

\authorinfo{Further author information: (Send correspondence to S.N. Zhang)\\Shuang-Nan Zhang: E-mail: zhangsn@ihep.ac.cn\\ }
\begin{document}
  \maketitle

%%%%%%%%%%%%%%%%%%%%%%%%%%%%%%%%%%%%%%%%%%%%%%%%%%%%%%%%%%%%%
\begin{abstract}
The High Energy cosmic-Radiation Detection (HERD) facility is one of several space astronomy payloads of the cosmic lighthouse program onboard China's Space Station, which is planned for operation starting around 2020 for about 10 years. The main scientific objectives of HERD are indirect dark matter search, precise cosmic ray spectrum and composition measurements up to the knee energy, and high energy gamma-ray monitoring and survey. HERD is composed of a 3-D cubic calorimeter (CALO) surrounded by microstrip silicon trackers (STKs) from five sides except the bottom. CALO is made of about 10$^4$ cubes of LYSO crystals, corresponding to about 55 radiation lengths and 3 nuclear interaction lengths, respectively. The top STK microstrips of seven X-Y layers are sandwiched with tungsten converters to make precise directional measurements of incoming electrons and gamma-rays. In the baseline design, each of the four side SKTs is made of only three layers microstrips. All STKs will also be used for measuring the charge and incoming directions of cosmic rays, as well as identifying back scattered tracks. With this design, HERD can achieve the following performance: energy resolution of 1\%  for electrons and gamma-rays beyond 100 GeV, 20\% for protons from 100 GeV to 1 PeV;  electron/proton separation power better than $10^{-5}$; effective geometrical factors of $>$3 ${\rm m}^{2}{\rm sr}$ for electron and diffuse gamma-rays, $>$2 $ {\rm m}^{2}{\rm sr}$ for cosmic ray nuclei. R\&D is under way for reading out the LYSO signals with optical fiber coupled to image intensified CCD and the prototype of one layer of CALO.
\end{abstract}

%>>>> Include a list of keywords after the abstract


%%%%%%%%%%%%%%%%%%%%%%%%%%%%%%%%%%%%%%%%%%%%%%%%%%%%%%%%%%%%%
\section{INTRODUCTION}
\label{sec:intro}  % \label{} allows reference to this section

%The China's space station is included in the space program directed by the China National Space Administration. It is a planned space station to be placed in low earth orbit (LEO) and to be launched around 2020.
It is well established that neutral, cold/warm and non-baryonic dark matter (DM) dominates the total matter content in the universe. Weakly Interacting Massive Particles\cite{wimp} (WIMPs) are well motivated candidates of DM particles. One way to detect WIMPs is to search for anomalies and sharp features in the observed spectra of cosmic electrons and gamma-rays. Some circumstantial evidence or hints of anomalies have been reported\cite{atic_electron,Fermi_electron}; however, astrophysical sources like pulsars and pulsar wind nebulae can also contribute to these results. Experimental data from more precise measurements at higher energies are still needed.

The steepening of the primary cosmic ray (CR) spectrum around several PeV, the so-called ``knee" structure is a classic problem in CR physics since its discovery in 1958, but still unresolved\cite{cr_model}. Ground-based extensive air shower experiments\cite{kascade,tibet,argo} continue to make progress \cite{ground_review}; however, these experiments have difficulties in making composition-resolved high-energy resolution measurements of the fine structure of the ``knee". On the other hand, experiments based on balloons\cite{atic2,cream1}, satellites\cite{pamela}, or the international space station\cite{ams02} can measure the particle energy and charge directly; however, these experiments suffer from small geometrical factor and limited energy range to make statistically meaningful measurements of the ``knee".

Several generations of wide field of view (FOV) space gamma-ray telescopes in the GeV energy regime and ground based narrow FOV gamma-ray telescopes in hundreds of GeV energy regimes have discovered several new populations of extreme astrophysical objects, which allow deeper understanding of the laws of nature under extreme physical conditions only available in cosmic laboratories. In particular the wide FOV space gamma-ray telescopes often provide crucial guidance to the observations of the ground-based narrow FOV telescopes sensitive. Unfortunately, the much more powerful ground-based Cherenkov Telescope Array (CTA) currently under development may not have the much needed guidance from a space wide FOV gamma-ray telescope, once the Fermi satellite stops operations. A new wide FOV space gamma-ray telescope is urgently needed to replace Fermi.

In order to address the above major problems in fundamental physics and astrophysics, the High Energy cosmic-Radiation Detection (HERD) facility has been planned as one of several space astronomy payloads of the cosmic lighthouse program onboard China's space station, which is planned for operation starting around 2020 for about 10 years. In this paper, we describe the scientific drivers of the design of HERD, its basic characteristics determined with Monte-Carlo simulations, as well as ongoing R\&D efforts in developing HERD.
%It's main scientific objectives are indirect dark matter search, precise measurements of cosmic ray spectra and composition up to the knee?energy, and high energy gamma-ray monitoring and survey.
%%%%%%%%%%%%%%%%%%%%%%%%%%%%%%%%%%%%%%%%%%%%%%%%%%%%%%%%%%%%%
\section{HERD Scientific Objectives, Requirements and Baseline Design}
The primary scientific objectives of HERD are: (1) searching for signatures of the annihilation products of dark matter particles in the energy spectra and anisotropy of high energy electrons and gamma-rays from 100 MeV to 10 TeV; (2) measuring precisely and directly the energy spectra and composition of primary cosmic rays from 10 GeV up to PeV. The secondary scientific objectives of HERD include wide FOV monitoring of the high energy gamma-ray sky from 100 MeV up to 10 TeV for gamma-ray bursts, active galactic nuclei and Galactic microquasars. Since models of dark matter particle annihilations do not yet have strong predictive power, our strategy in the baseline design of HERD is to ensure that the effective geometrical factor, energy range and resolution of HERD meet the requirements for observations of cosmic rays, while maintaining the best possible capability in observing electrons and gamma-rays within the currently available resources for placing HERD on board China's space station. Extensions to the baseline design may be made to increase its effective geometrical factor for gamma-rays (and electrons) with excellent energy and angular resolution.

\begin{figure}
\begin{center}
\begin{tabular}{c}
\hbox{\includegraphics[height=6cm]{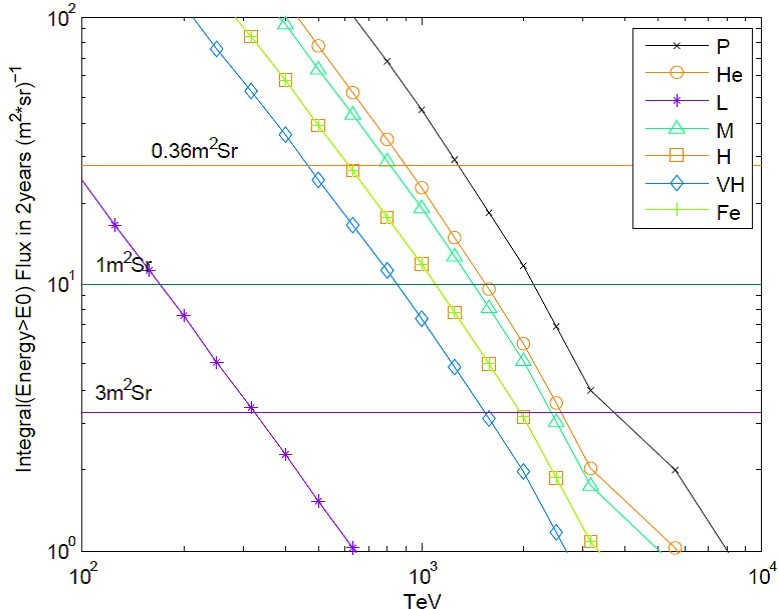}\hspace{0.5cm}\includegraphics[height=6cm]{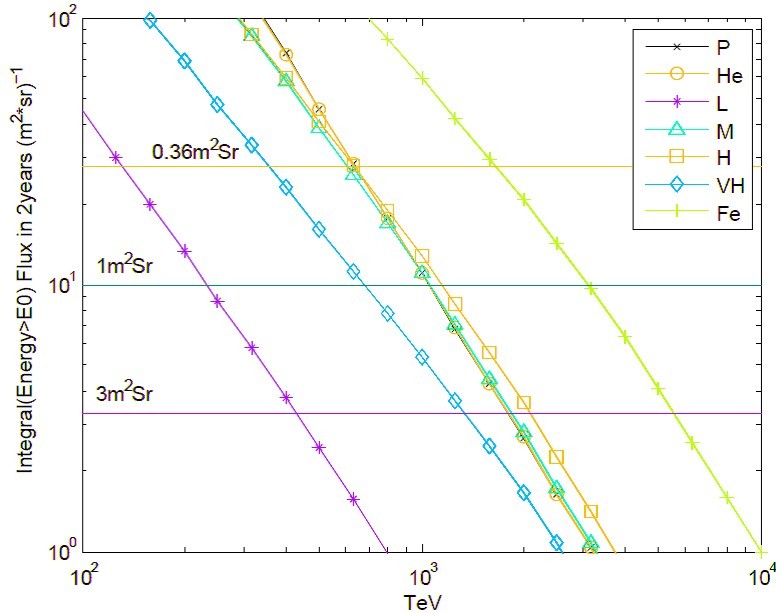}}
\end{tabular}
\end{center}
\caption[example]
%>>>> use \label inside caption to get Fig. number with \ref{}
{ \label{fig:requirement}
Total number of events per m$^2$sr in two years as a function of threshold energy for each of the cosmic ray compositions, predicted with the hard and nonlinear acceleration models\cite{tibet}. The horizontal lines show the flux required above a certain energy for detecting 10 events in two years with different effective geometrical factor in units of m$^2$sr. In the left and right panels, cosmic-rays are dominated by protons and iron nuclei, respectively. For convenience we define several groups of elements: ``L" with $3\le Z \le 5$, ``M" with $6\le Z \le 9$, ``H" with $10\le Z \le 19$, ``VH"  with $Z \ge 20$ but excluding Fe.}
\end{figure}

In Fig.~\ref{fig:requirement}, we show the model predicted fluxes for different compositions of cosmic rays with the hard and nonlinear acceleration models\cite{tibet}. We considered two extreme cases, i.e., cosmic rays are dominated by protons or iron nuclei, respectively. We conclude that a geometrical factor (with 100\% efficiency) of $\sim$3~m$^2$sr is required in order to detect at least 10 events above PeV for all groups of nuclei, except the very rare ``L" group with $3\le Z \le 5$, i.e. Lithium, Beryllium and Boron. Our design goal for the calorimeter of HERD is thus simply to achieve an effective geometrical factor of $\sim$3~m$^2$sr after taking into account the detection and event reconstruction efficiency. To do this, we find that the HERD baseline design with a cubic calorimeter (CALO) of 63~cm$\times 63$~cm$\times 63$~cm is required, which is made of nearly 10$^4$ pieces of granulated LYSO crystals of 3~cm$\times 3$~cm$\times 3$~cm each. From any incident directions, CALO has a minimum stopping power of $55X_0$ and $3\lambda$, where $X_0$ and $\lambda$ are radiation and nuclear interaction lengths, respectively. Such a deep and high granularity calorimeter is also essential for excellent electron-proton separation and energy resolutions of all particles. It also has some directional measurement capability with the reconstructed 3-D showers.

\begin{figure}
\begin{center}
\begin{tabular}{c}
\hbox{\includegraphics[height=7cm]{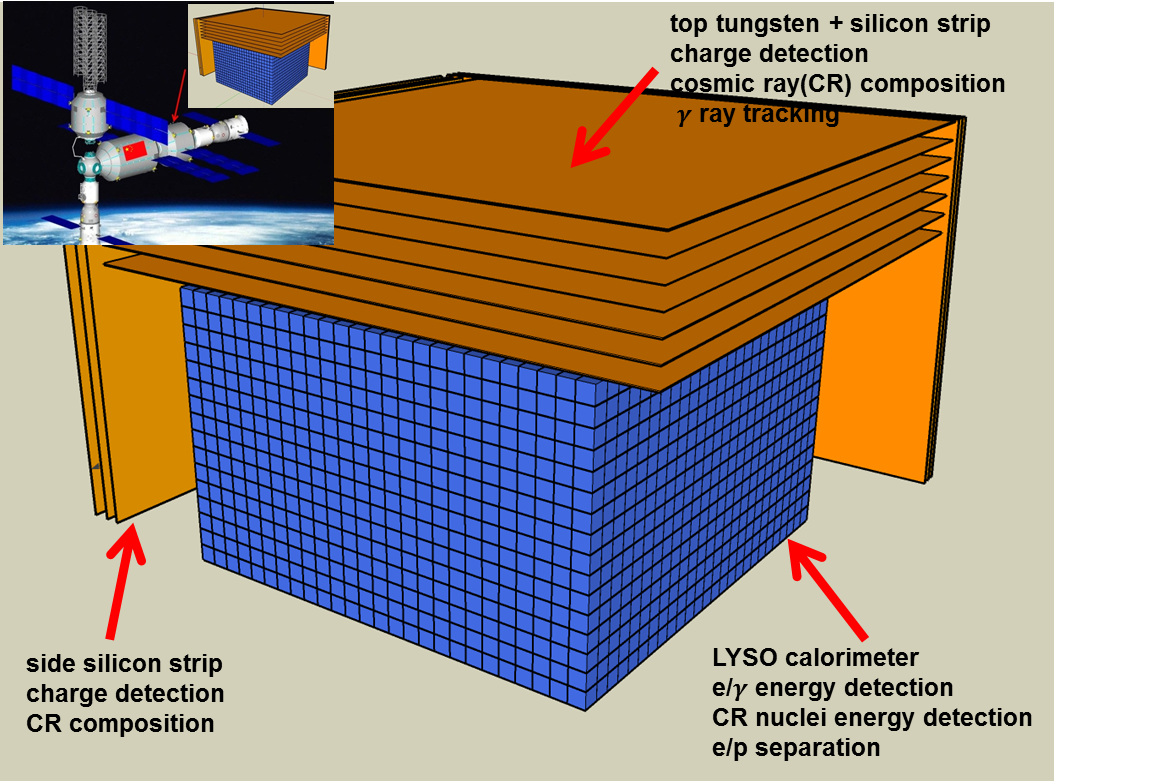}\hspace{1cm}\includegraphics[height=7cm]{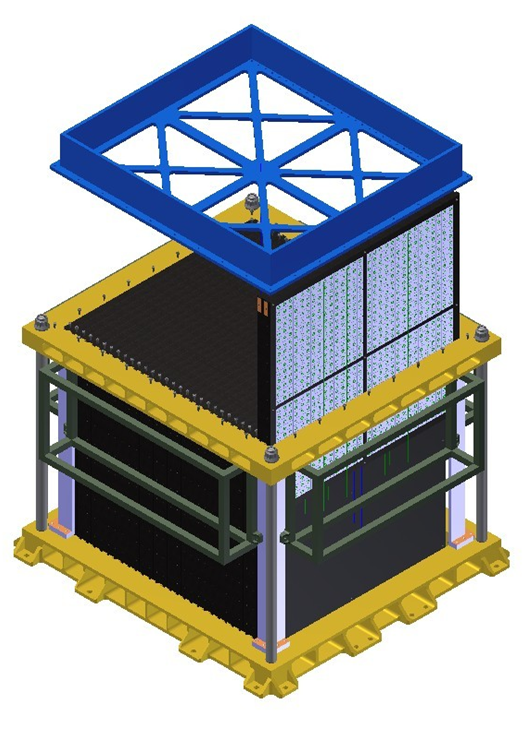}}
\end{tabular}
\end{center}
\caption[example]
%>>>> use \label inside caption to get Fig. number with \ref{}
{ \label{fig:baseline}
{\it Left}: Schematic diagram of the baseline design of HERD. The top STK is made of seven layers of silicon microstrips, sandwiched with tungsten foils; however the STKs on the four side are made of only three layers of silicon microstrips without tungsten foils. The extended design of HERD will have its four-side STKs replaced by STKs almost identical to the top STK. {\it Right}: Prototype of HERD calorimeter.}
\end{figure}

In order to measure the charges and incident directions of cosmic rays, silicon trackers (STKs) are required with a minimum of three layers of silicon micro-strip detectors (SSDs), which can also be used to reject backslash tracks from the showers in CALO. To measure accurately the incident directions of  gamma-rays, electron-position pairs should be created and tracked; this can be achieved by adding tungsten foils as shower converters and four more layers of SSDs. In the baseline design of HERD, only the top STK is equipped with seven layers of SSDs sandwiched with tungsten foils, as shown in Fig.~\ref{fig:baseline} (left); the right panel is an illustration for a laboratory prototype of CALO. A possible option, as an extended design of HERD, is to surround CALO by the same seven-layer STK with tungsten foils from all four sides, to ensure the maximum FOV for electrons and gamma-rays. Plastic scintillators surrounding HERD from all five sides may be needed to reject most low energy charged particles, in order to have maximum efficiency for high energy cosmic rays and  electrons, as well as gamma-rays of all energies. The HERD baseline characteristics and main functions of its CALO and STKs are listed in Table 1.

\begin{table}
\begin{center}
 \caption{HERD baseline characteristics of all components.}
 \begin{tabular}{cccccc}
  \hline \hline
  &type &size &$X_0$,$\lambda$ &unit &main functions\\
 \hline
 Top STK &Si strips &70 $\times$ 70 cm$^2$ &2 $X_0$ &7 x-y (W foils) &Charge, Early shower, Tracks\\
4-side STK &Si strips &65 $\times$ 50 cm$^2$ &-- &3 x-y &Nucleon Tracks, Charge\\
CALO & $\sim10^4$ LYSO &63 $\times$ 63 &55 $X_0$, 3 $\lambda$ &3 $\times$ 3 &e/$\gamma$ energy, nucleon energy,\\
& cubes & $\times$ 63 cm$^3$ & &$\times$ 3 cm$^3$  &e/p separation\\
  \hline \hline
 \end{tabular}
 \end{center}
\label{baseline_table}
\end{table}

\section{Expected Performance of the HERD Baseline Design}
\label{sec:simu}
Extensive simulations have been carried out with GEANT4\cite{geant4} and FLUKA\cite{fluka_1}, in order to evaluate the scientific performance of the HERD baseline design  and to optimize the relative weights of each component of HERD within the boundary conditions for accommodating HERD on board China's space station. Since the performance of CALO is key to meet the scientific goals of HERD, here we only present our simulation results of CALO, by focusing on its effective geometrical factor, energy resolution and e/p separation capability, in order to predict the observed cosmic ray spectra. A key assumption is that an average of 10 photoelectrons can be collected per minimum ionization particle response, which is the design goal of our readout system and already demonstrated in our laboratory test system. For its sensitivity of gamma-ray continuum all sky survey and line observations, certain assumptions are made for its STKs, based on primarily the performance of SSDs of Fermi and AMS02. To simplify the simulations, no mechanical and other supporting structures and materials are included in the simulations. In Table 2, we list the expected HERD baseline performance from Monte-Carlo simulations; please refer to Xu et al (2014) in the same proceedings for further details of the simulations\cite{XuMing}.

\begin{figure}
\begin{center}
\begin{tabular}{c}
\includegraphics[width=13cm]{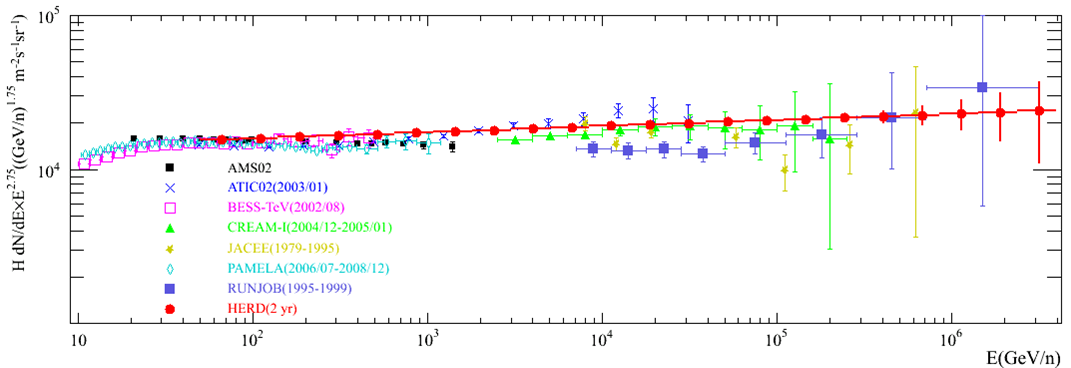}\\
\includegraphics[width=13cm]{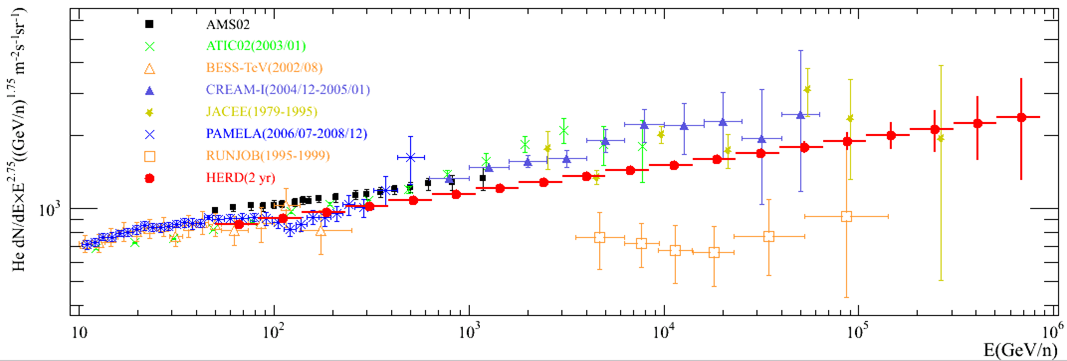}\\
\includegraphics[width=13cm]{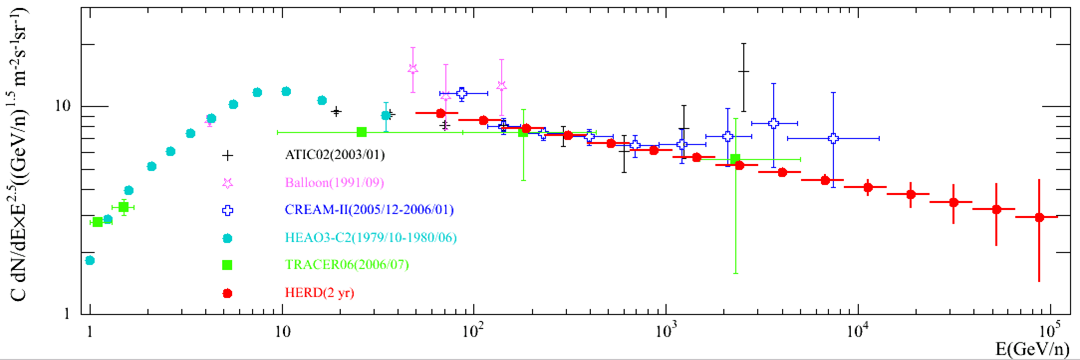}
\\\includegraphics[width=13cm]{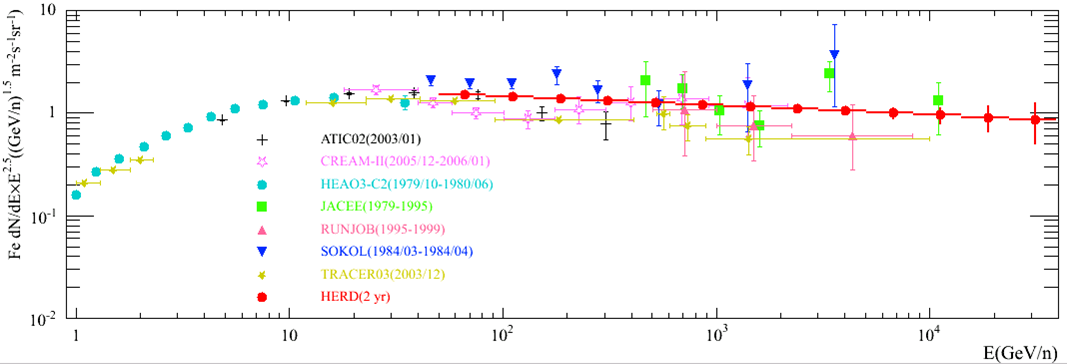}
\end{tabular}
\end{center}
\caption[example]
%>>>> use \label inside caption to get Fig. number with \ref{}
{ \label{fig:crspectra}
Simulated two-year HERD cosmic ray spectra of protons, heliums, carbons and irons (from top to bottom), in comparison with previous direct measurements in space or at balloon altitudes. The input cosmic ray composition model\cite{horandel_model} for the simulation is a combined fitted result from previous measurements.}
\end{figure}

The key performance of HERD, in comparison with all previous and other approved missions, is its extremely large effective geometrical factor for all types of high energy cosmic radiations, thanks to its very deep 3-D CALO and five-side STKs. In Fig.~\ref{fig:crspectra}, we show the predicted HERD spectra for protons, helium nuclei, carbon nuclei and iron nuclei, in comparison with all previous direct measurements in space. Clearly HERD will surpass all previous results of directly measured cosmic rays from, e.g., AMS02\cite{ams02}, ATIC-2\cite{atic2}, BESS\cite{bess}, CREAM,\cite{cream1,cream2} HEAO\cite{heao}, JACEE\cite{jacee}, PAMELA\cite{pamela}, RUNJOB\cite{runjob}, SOKOL\cite{sokol} and TRACER\cite{tracer}, with much better statistics and up to much higher energies even beyond PeV and into the ``knee" region. For example, at least ten events will be recorded from 900 TeV to 2 PeV for each specie, which means that the energy spectra of most nuclei will be directly extended to the knee range with much smaller error bars than previous direct measurements in space.

\begin{table}
\begin{center}
 \caption{HERD baseline performance. Note that in the baseline design, i.e., only the top STK has seven layers of SSDs sandwiched with tungsten foils, which is expected to deliver an angular resolution of 0.1$^\circ$ with $\sim 1$ m$^2$sr. In the extended HERD design, all five sides have almost identical SKTs as the top STK. It should also be noted that the current HERD STKs can deliver only very poor angular resolution down to the 100 MeV lower energy limit for gamma-rays; further significant improvements in the STK design are required to enhance its low energy gamma-ray capability.}
 \begin{tabular}{cccccccccc}
  \hline \hline
 Particle&$\gamma$/e &CR &$\gamma$/e &CR &$\gamma$/e  &p &e/p &e&p \\
Para.&energy &energy &$\Delta\theta$&$\Delta C$&$\Delta E/E$&$\Delta E/E$&sep. &$A_{\rm eff}$&$A_{\rm eff}$\\
Perfor.&100 MeV & GeV &0.1$^\circ$&~$\sim$0.1 c.u& $<$1\% &20\%&$<10^{-5}$&3.7 m$^2$sr&2.6 m$^2$sr \\
           &-10 TeV & -PeV &&& @200 GeV& &&@600 GeV& @400 TeV \\
Detector&CALO&CALO&T-STK&
All STK&CALO&CALO&CALO&CALO&CALO\\
  \hline \hline
 \end{tabular}
 \end{center}
\label{performance}
\end{table}

\begin{figure}
\begin{center}
\begin{tabular}{c}
\hbox{\includegraphics[height=5cm]{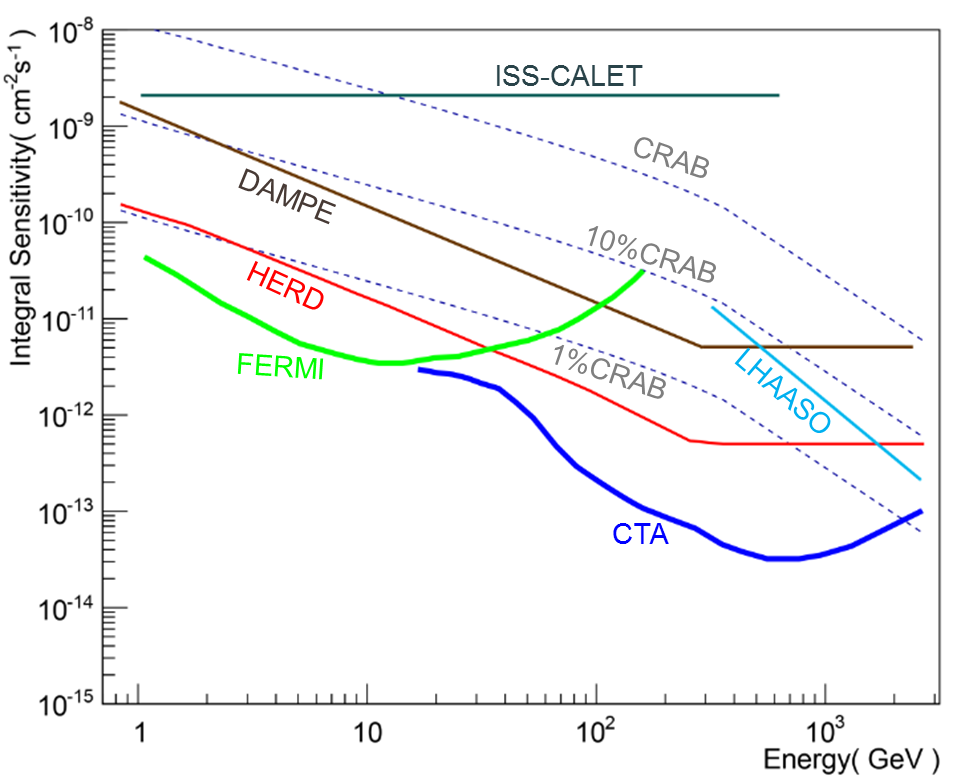}\hspace{1cm}\includegraphics[height=5.5cm]{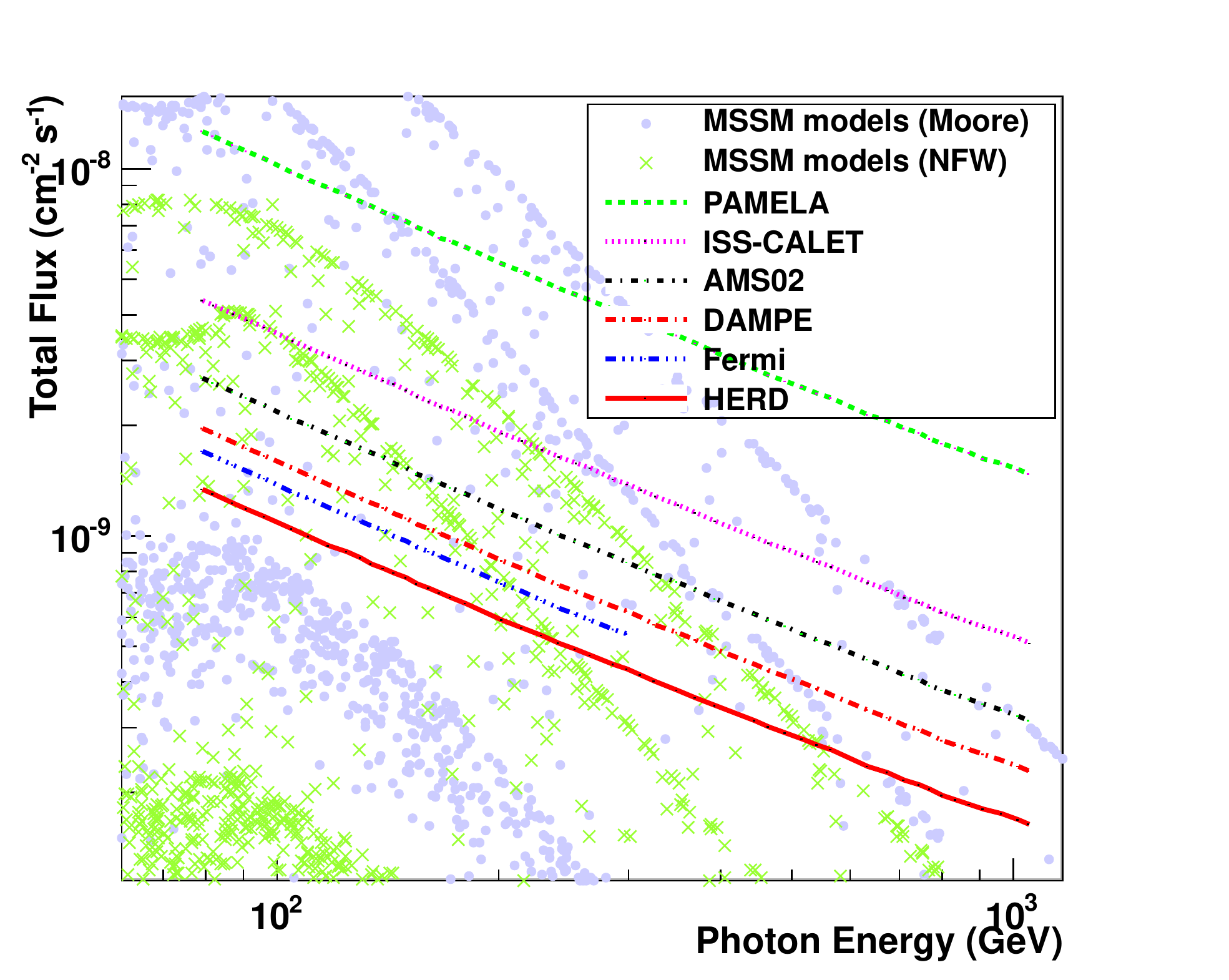}}
\end{tabular}
\end{center}
\caption[example]
%>>>> use \label inside caption to get Fig. number with \ref{}
{ \label{fig:gammaray}
Expected gamma-ray sky survey sensitivity of the extended HERD design, i.e., all five sides are surrounded by the same seven-layer STKs with 0.1$^\circ$ angular resolution across the whole energy band; for the baseline design, the sensitivity is degraded by nearly a factor of 2. {\it Left}: HERD $5\sigma$ continuum sensitivity for one year observation in comparison with all other missions with gamma-ray observation capability, e.g., ISS-CALET\cite{CALET_crab}, DAMPE and Fermi\cite{Fermi_crab}, and including the future ground based Cherenkov Telescope Array (CTA)\cite{CTA_crab} high energy gamma-ray telescope and the Large High Altitude Air Shower Observatory (LHAASO)\cite{LHAASO_crab}. {\it Right}: HERD one-year $5\sigma$ line sensitivity in comparison with predictions of different dark matter models; the sensitivity lines of other experiments are calculated with the following operation periods: 2006-2016 (PAMELA), 2016-2021 (CALET); 2011-2021 (AMS02), 2016-2021 (DAMPE), 2008-2018 (Fermi).}
\end{figure}

With an adequate design of STKs, HERD will also have adequate capability for gamma-ray observations, as shown in Fig.~\ref{fig:gammaray}. In the post-Fermi era, HERD will be the most sensitive gamma-ray all-sky survey and transient monitor from GeV to around TeV, an essential capability to provide triggers and alerts to other multi-wavelength telescopes, such as the future ground based Cherenkov Telescope Array (CTA) high energy gamma-ray telescope. It is widely anticipated that  gamma ray emission lines are the smoking guns for identifying dark matter particle annihilations. As shown in the right panel of Fig.~\ref{fig:gammaray}, HERD's line sensitivity is far superior to all other missions, due to the combination of its excellent energy resolution, very large effective geometrical factor and high background rejection efficiency.

%\cite{HESS_crab_nebula, Fermi_crab_nebula}

\section{R\&D of HERD Calorimeter}

\begin{figure}
\begin{center}
\begin{tabular}{c}
\includegraphics[height=7cm]{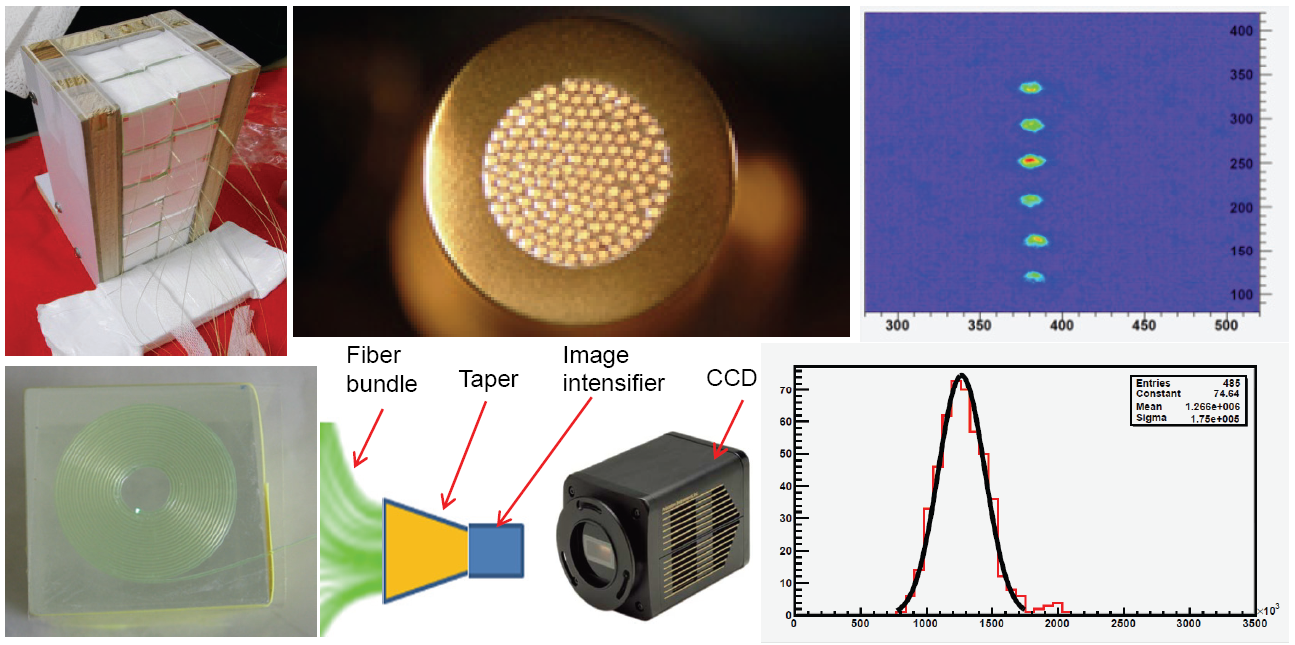}
\end{tabular}
\end{center}
\caption[example]
%>>>> use \label inside caption to get Fig. number with \ref{}
{ \label{fig:labresults}
Laboratory test result with $2\times2\times6$ CsI (Na) crystal cubes coupled by optical fibers, wavelength shifters, a taper, an image intensifier and a CCD. The signals are from cosmic ray muons.}
\end{figure}

\begin{figure}
\begin{center}
\begin{tabular}{c}
\hbox{\includegraphics[height=5cm]{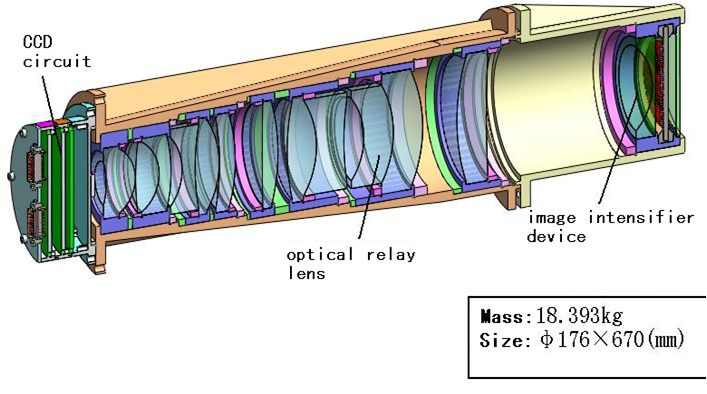}\hspace{1cm}\includegraphics[height=5cm]{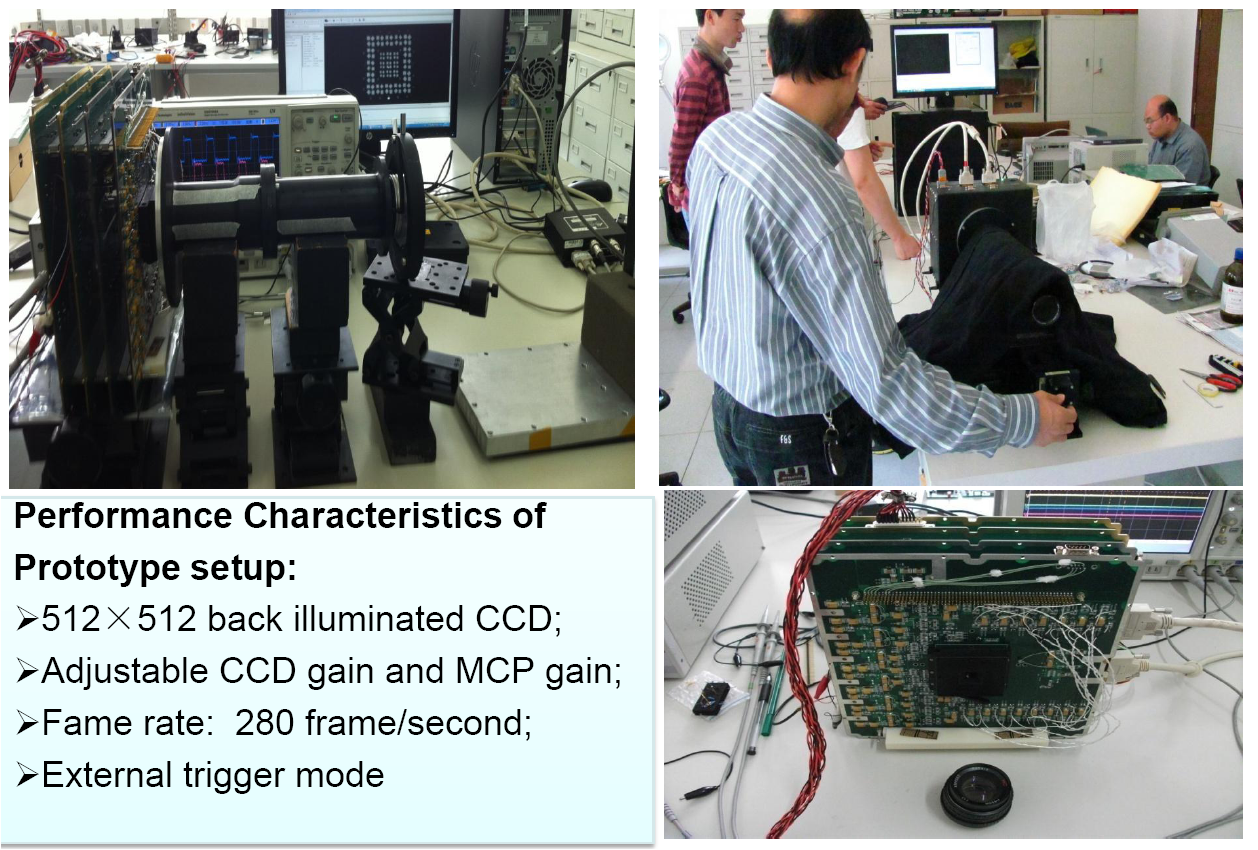}}\\
\hbox{\includegraphics[height=5cm]{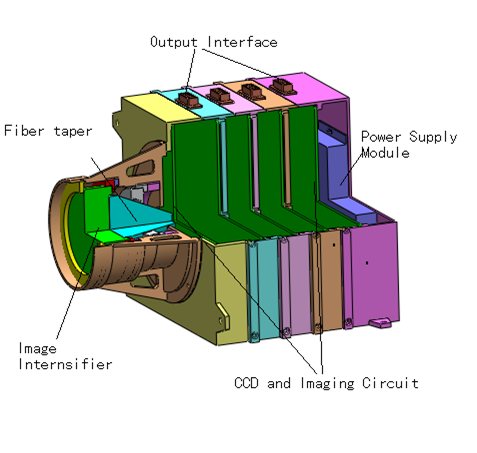}\hspace{1.3cm}\includegraphics[height=5cm]{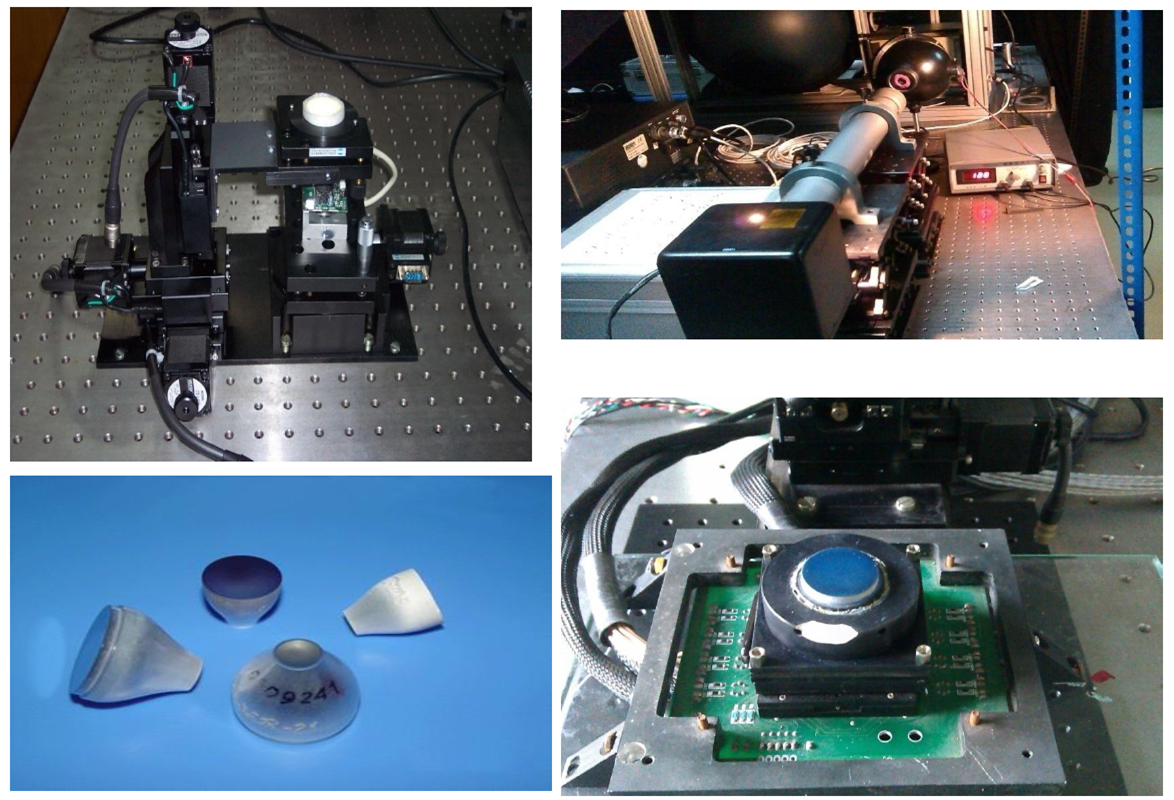}}
\end{tabular}
\end{center}
\caption[example]
%>>>> use \label inside caption to get Fig. number with \ref{}
{ \label{fig:coupling}
Optical couplings between readout optical fibers and CCD. {\it Top}: schematic design and experimental setup of a relay lens system with an image intensified CCD. {\it Bottom}: schematic design and experimental setup of a taper system with an image intensified CCD.}
\end{figure}

A key technology of the HERD mission is the signal readout system of the $10^4$ pieces of LYSO crystals with sufficient signal to noise ratio and large dynamical range. In order to minimize the power consumption of readout electronics and heat dissipation inside CALO, we choose to channel the scintillation light out of the LYSO crystals with optical coupling onto CCDs, as illustrated in Fig.~\ref{fig:labresults} with laboratory test results\cite{sun}. To demonstrate the feasibility of applying such a signal readout system in space, we are currently developing some engineering models of complete readout systems. Two approaches of the optical couplings are investigated: a relay lens system or optical taper between the fibers and an image intensifier, as shown in Fig.~\ref{fig:coupling}. The advantage of the relay lens system is that there is no hard contact to the fibers and image intensifier; however, transmission efficiency is expected to be very low. The pro and cons of the taper approach are just opposite.

Our simulations require: (1) The dynamical range of at least $2\times 10^6$ are required, because high energy cosmic rays often produce large numbers of secondary particles passing through a single LYSO crystal in the early shower development; (2) An end-to-end efficiency, resulting in a minimum of 10 photoelectrons produced from the cathode of the image intensifier for a minimum ionization energy response, is required, in order to achieve the best possible energy resolution.

To address the dynamical range requirement, a beam test is on-going to evaluate if such a cube LYSO crystal can respond to ionization energy deposition over such a dynamical range; finer crystals can be used to reduce the dynamical range requirement. However, neither image intensifiers nor CCDs can handle such a huge dynamical range. Therefore we need to split the light output of each crystal into two channels by a ratio of 1:1000, which then requires a dynamical range of only slightly more than 1000 for the image intensifiers and CCDs.

Our current results indicate that the 10-photoelectron requirement can be met with the taper coupling approach, but very difficult with the relay lens system. Since each shower lights up several hundreds crystals, a higher efficiency improves the energy resolution only marginally. Actually, we anticipate that the systematic errors in the gain or response calibrations of the crystals will dominate the eventual energy resolution at system level, which can be realistically assessed through laboratory tests. It is therefore essential to build a portion of CALO to allow end-to-end performance evaluation of CALO in high energy hadron beam tests, including event reconstruction algorithms; our estimate is that 1/20 of the full CALO in a long cylindrical form is sufficient.

%%%%%%%%%%%%%%%%%%%%%%%%%%%%%%%%%%%%%%%%%%%%%%%%%%%%%%%%%%%%%
\acknowledgments     %>>>> equivalent to \section*{ACKNOWLEDGMENTS}
The authors would like to thank funding supports from the Chinese Strategic Pioneer Program in Space Science under Grant No.XDA04075600, the Qianren start-up grant 292012312D1117210, National Natural Science Foundation of China under Grant No.11327303, and the Cross-disciplinary Collaborative Teams Program for Science, Technology and Innovation, Chinese Academy of Sciences (Research Team of The High Energy cosmic-Radiation Detection).
%%%%%%%%%%%%%%%%%%%%%%%%%%%%%%%%%%%%%%%%%%%%%%%%%%%%%%%%%%%%
%%%%% References %%%%%

\bibliography{report}   %>>>> bibliography data in report.bib
\bibliographystyle{spiebib}   %>>>> makes bibtex use spiebib.bst

\end{document}